\documentclass{appolb}
\usepackage{epsfig}

\begin{document}

\title{GDR Feeding of the Highly-Deformed Band in $^{42}$Ca
\thanks{
       Presented at the XXXIX Zakopane School of Physics - International 
       Symposium "Atomic nuclei at extreme values of temperature, spin and 
       isospin" - Zakopane, Poland, August 31 - September 5, 2004}
}

\author{M.~Kmiecik$^a$, A.~Maj$^a$, J.~Stycze\'n$^a$, P.~Bednarczyk$^{a,d,e}$, 
        M.~Brekiesz$^a$, J.~Gr\c{e}bosz$^a$, M.~Lach$^a$, W.~M\c{e}czy\'nski$^a$, 
        M.~Zi\c{e}bli\'nski$^a$, K.~Zuber$^a$, A.~Bracco$^b$, F.~Camera$^b$, 
	G.~Benzoni$^b$, B.~Million$^b$, S.~Leoni$^b$, O.~Wieland$^b$, 
	B.~Herskind$^c$, D.~Curien$^d$,\\  
	N.~Dubray$^d$, J.~Dudek$^d$, N.~Schunck$^c$ and K.~Mazurek$^{f,a}$
\address{$^a$The Niewodnicza\'nski Institute of Nuclear Physics, PAN, Krak\'ow, Poland \\
$^b$University of Milano - Department of Physics and INFN section of Milano, Italy\\
$^c$The Niels Bohr Insitute, Copenhagen, Denmark \\
$^d$IReS IN$_{2}$P$_{3}$-CNRS/Universit\'e Louis Pasteur, Strasbourg, France\\ 
$^e$Gesellschaft f\"ur Schwerionenforschung, Darmstadt, Germany\\
$^f$Katedra Fizyki Teoretycznej, UMCS, Lublin, Poland}
}

\maketitle

\begin{abstract}

The $\gamma$-ray spectra from the decay of the GDR in the compound nucleus
reaction $^{18}$O+$^{28}$Si at bombarding energy of 105~MeV have been measured
in an experiment using the EUROBALL IV and HECTOR arrays. The obtained
experimental GDR strength function is highly fragmented, with a low energy
($\approx$10~MeV) component, indicating a presence of a large deformation and Coriolis
effects. In addition, the preferential feeding of the highly-deformed band in
$^{42}$Ca by this GDR low energy component is observed.

\end{abstract}

  
\section{Introduction}
\label{Sect.01}

Change in nuclear shape from an oblate one with the spin parallel to the
symmetry axis to an elongated prolate or triaxial one, accompanied by the
collective rotation around the shortest axis, called Jacobi shape transition, 
has been predicted to appear in many nuclei at angular momenta close to the
fission limit. The presence of elongated shapes has been indicated in the
$\gamma$-decay of the  Giant Dipole Resonance (GDR) in $^{46}$Ti
nucleus~\cite{Maj_NS2002,Maj04}. In this article we show further observations, which
confirm this finding. The simulation of the GDR strength function based on the
Lublin-Strasbourg Drop
(LSD) model~\cite{Pom03,Dud04,Dub04}, when inserted to the statistical 
evaporation code CASCADE, results
in a high-energy $\gamma$-spectrum very similar to the experimental one. The
low energy component of the GDR strength function is interpreted as a result of
both large deformation effects and Coriolis splitting of the GDR. Moreover, the
low energy component of the GDR has been found to feed preferentially the
highly-deformed band~\cite{Lach} in $^{42}$Ca.


\section{Jacobi shape transitions and Coriolis splitting of the GDR}
\label{Sect.02}

The experiment~\cite{Maj_NS2002,Maj04} was performed at the VIVITRON accelerator of the
IReS Laboratory, in Strasbourg (France), using the EUROBALL~IV
Ge-array~\cite{Beck,Simps} coupled  to the BaF$_2$ HECTOR array~\cite{Maj94}. 
The master trigger
condition was highly selective - accepted were only the events having at  least two
Ge signals of EUROBALL, BGO-shield suppressed,  and one high-energy
$\gamma$-ray in BaF$_2$ detector of HECTOR.  The $^{46}$Ti compound nucleus was
populated in the $^{18}$O+$^{28}$Si reaction at 105~MeV bombarding energy. The
excitation energy of the $^{46}$Ti nuclei  was 86~MeV and the maximum angular
momentum, $\ell_{max}\approx$~35~$\hbar$. 

The GDR spectrum measured by HECTOR detectors, gated on known well resolved 
low energy $\gamma$-ray transitions of $^{42}$Ca detected in EUROBALL, is shown in Fig.~1a.
This gating condition, together with the highly selective master trigger condition,
allowed to select high energy $\gamma$-rays coming from nuclei with the highest 
spins ($>$~20~$\hbar$) and free from fission and direct reactions 
contaminations.

The high-energy $\gamma$-ray spectrum was analysed using a modified Monte-Carlo
version of the statistical model code CASCADE~\cite{Her,Puhl}. 
The GDR line shape given by the
absorption cross-section extracted using the method described in e.g.
Ref.~\cite{Kic93} is shown in the inset to Fig.~1a. The found strength function is split
forming a narrow low-energy component around 10~MeV, and a broad structure
ranging from 15 to 27~MeV. This splitting was interpreted~\cite{Maj04} as the
consequence of both the Jacobi shape transition in which an oblate nucleus,
with a non-collective rotation transforms (around spin $I$~=~28~$\hbar$) to a
very elongated  prolate one rotating collectively, and strong
Coriolis effects which split the low energy component (at $\approx$~13 MeV)
by $2 \omega_{rot}$~$\approx$~6~MeV (where $\omega_{rot}$ is the rotational frequency)
and shift a part of its strength down to $\approx$10~MeV. In fact, Ref.~\cite{Maj04}
presented the first clear observation of the Coriolis effects in hot nuclei. 

\begin{figure}[htb]
\begin{center}
\epsfig{file=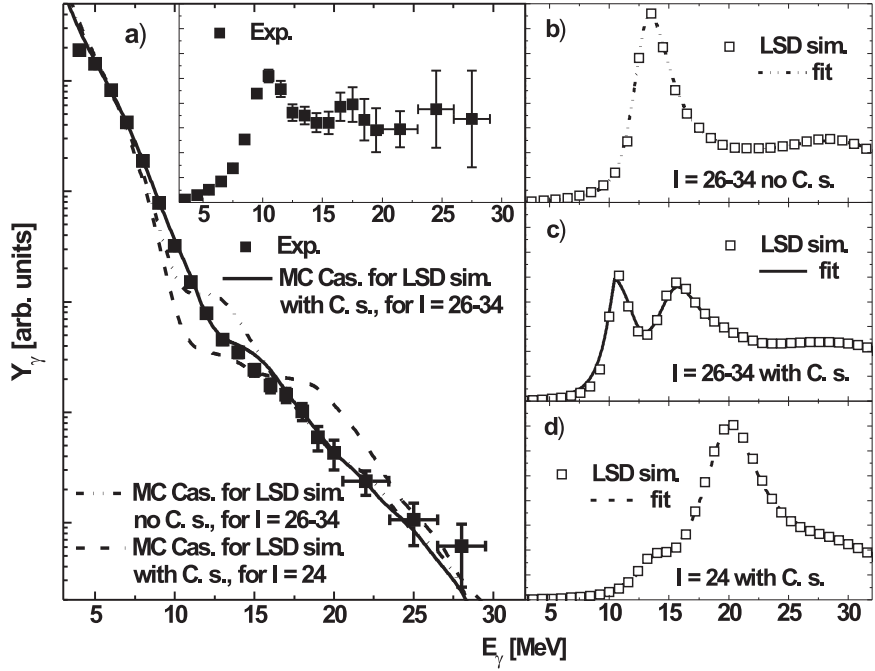,height=9.0cm}
\caption{\textbf{a)} Full squares represent the experimental high energy 
         $\gamma$-ray spectrum~\cite{Maj_NS2002,Maj04} 
	  compared to the spectra calculated using Monte Carlo Cascade code 
         for GDR parameters obtained from the fit to the LSD model simulation~\cite{Dub04} for 
         spins indicated. In the inset the experimental GDR strength function is shown.
         Lines show the results of the CASCADE calculations with the GDR parameters fitted to the
	  theoretical predictions shown in panels a), b) and c);
         \textbf{b)}~Open squares represent theoretical simulation results 
         (in the spin range from 26 to 34~$\hbar$) of the GDR 
         line shape in $^{46}$Ti, obtained from the thermal shape fluctuation model based 
         on free energies from the LSD model calculations, in which the Coriolis 
         splitting of the GDR components 
         is not included. The dash-dotted line is the fit of simulated GDR
         line shape with a 5-component Lorentz function;
         \textbf{c)} The LSD simulation results for the same spin range but including
         the Coriolis splitting of the GDR components (points) with the fit
         presented with solid line;
         \textbf{d)} Similar to c) but for $I$=~24~$\hbar$. The dashed line shows 
         the fit of simulated GDR line shape.}
\end{center}
\end{figure}

In order to confirm this interpretation, in the following we compare the
experimental spectrum with the calculated spectra in which the GDR strength
function is simulated according to the theoretical predictions of the nuclear
shape. The simulations are done using the method of the thermal shape
fluctuations (viz. Ref.~\cite{Ormand,JJG} and references therein) 
based on the free energies from the newest version of the liquid
drop model LSD~\cite{Pom03}, and on the microscopic single-particle spectra from
the deformed Woods-Saxon Hamiltonian with the {\em universal} parameters. In the
calculations~\cite{Maj04,Dub04}, we include the possibility of Coriolis
splitting of the GDR strength function for a given spin and a deformation
value. Thus, the GDR line shape consists,  in general, of 5-Lorentzian
parametrisation~\cite{Nee82,Gall,Doss}.

The results of the simulation for the spin region $I$~$>$~26$~\hbar$, 
where the Jacobi transition
is predicted to occur, are presented as open squares in Fig.~1b,~1c, and denoted as
"LSD sim". In Fig.~1b the Coriolis splitting of the GDR components was not included
while it was done for calculations shown in Fig.~1c. The latter are indeed similar 
to the experimental results for the GDR
strength function (Fig.~1a - inset), and also exhibit the narrow 10~MeV component  
and a broad structure at higher energies. For comparison, the calculated GDR line
shape for $I$~=~24~$\hbar$, i.e. in the region where the small oblate deformation 
is expected, is shown in Fig.~1d. 

The simulated GDR strength function can be used in the statistical model
calculations. We have modified the Monte-Carlo version of the code CASCADE, so
that the GDR strength function  is parameterised by 5 components, each defined
by a GDR centroid, a width and a strength. Those 15 input parameters have been
obtained by fitting the simulated GDR strength functions (Fig.~1b,~c,~d) with a
5-component Lorentz function -  which is shown in Fig.~1b, 1c and 1d by dash-dotted, 
solid and broken lines, respectively.

The theoretical spectra resulting from the CASCADE calculations are shown also
in Fig.~1a  together with the experimental data. As can be seen, the
calculations done for the spin range 26-34~$\hbar$ taking into account the Coriolis
splitting of the GDR components are in very good agreement with the experimental
spectrum,  while the one without Coriolis splitting assumption 
or the one for $I$~=~24~$\hbar$ are in full disagreement. This gives the
additional confirmation of the observation of the Jacobi shape transition in
the hot $^{46}$Ti and the Coriolis splitting of the GDR.

In this context it is worth to mention, that very large deformations of $^{46}$Ti
at high angular momenta
were also suggested by the measured spectra of emitted $\alpha$-particles~\cite{Brekiesz}.

\section{The GDR feeding of various bands in $^{42}$Ca}

In addition, the GDR feeding of the normal-deformed bands and the
highly-deformed  (possibly super-deformed) collective band in
$^{42}$Ca~\cite{Lach} was investigated. To see how the different regions of
high-energy $\gamma$-rays feed the discrete lines in $^{42}$Ca residual
nucleus, the gates (1~MeV wide) were set on the GDR spectrum and with such a
condition the discrete line intensities were analysed. The three ratios between
intensities of different discrete transition (normal deformed,  highly-deformed
and spherical) were taken into account. They are plotted as functions of the
GDR energy in Fig.~2 after an arbitrary normalisation to 1 at 4.5~MeV.

As clearly seen, the ratio between $\gamma$-intensities from the
highly-deformed (denoted as "hd") and normal-deformed ("nd") bands (solid
points) shows a bump in the region 8-9~MeV.  In this region the ratio is
larger by a factor of 2 as compared to the low energy (statistical) region, and
by a factor almost 4 as compared to  the normal GDR region ($>$12~MeV). An
enhancement, but smaller, is seen also in the ratio of the intensities of transitions
within the "hd"-band and in the spherical
("spher") part of the level scheme (open squares),
while there is no bump present in
the "nd/spher" ratio case (open triangles). Considering that the gates
were set on the raw spectrum, not corrected for the detector's  response
function, this 8-9 MeV bump corresponds to the 10~MeV low energy component 
of the GDR strength function shown in Fig.~1a. This might indicate that the low
energy component of the Coriolis split GDR  in the {\it hot} compound nucleus
$^{46}$Ti feeds preferentially the highly-deformed  (presumably super-deformed)
band in the {\it cold} $^{42}$Ca evaporation residue. Similar effect has been
observed in the $^{143}$Eu case~\cite{Benzoni}.  This seems to confirm the  old
hypothesis~\cite{Herskind} of a special role played by the low energy GDR 
component in feeding the super-deformed yrast structures. 

\begin{figure}[htb]
\begin{center}
\epsfig{file=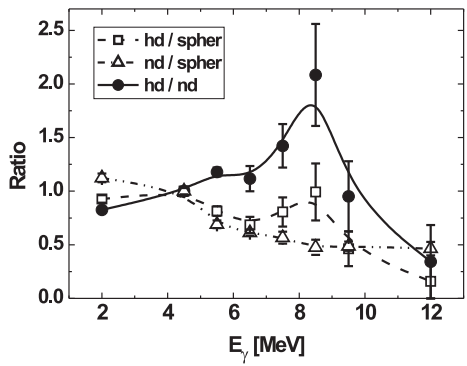,height=4.8cm}
\caption{The intensity ratios between gamma transitions proceeding within
         spherical (denoted as "spher"), normal deformed ("nd") and  
         highly-deformed ("hd") bands as a function of the $\gamma$-ray 
         energy from the GDR decay.}
\end{center}
\end{figure}

\section{Summary}

The high-energy $\gamma$-ray spectrum from the $^{46}$Ti compound nucleus
measured in coincidence with discrete transitions in the $^{42}$Ca residues
shows highly fragmented GDR strength function with a broad 15-25 MeV structure
and a narrow low-energy 10~MeV component. This can be interpreted as the
result of the Jacobi shape transition and strong Coriolis effects, 
seen experimentally for
the first time. In addition, the low energy GDR component seems to feed
preferentially the highly-deformed band in $^{42}$Ca. This suggests that the
very deformed shapes of hot compound nucleus, resulting from the Jacobi shape
transition, persist in the entire evaporation process. Thus, the mechanism of
the Jacobi shape transition might indeed constitute a gateway to the production
of the very elongated, rapidly rotating and relatively cold nuclei, as
advocated in Ref.~\cite{Dudek}.

\section{Acknowledgements}
This work was partially supported by the Polish Committee for Scientific
Research (KBN Grant No. 2 P03B 118 22), by the exchange programme between
the {\em Institut National de Physique Nucl\'eaire et de Physique des
Particules, $IN_2P_3$}, and Polish Nuclear Physics Laboratories,
and by the European Commission contract HPRI-CT-1999-00078.


\begin{thebibliography}{21}

\bibitem{Maj_NS2002} A. Maj et al.,
			{\it Eur. Phys. J.} \textbf {A20}, 165 (2004).
\bibitem{Maj04} A. Maj et al., 
               	{\it Nucl. Phys.} \textbf {A731c}, 319 (2004). 
\bibitem{Pom03} K. Pomorski and J. Dudek, 
               	{\it Phys. Rev.} \textbf {C67}, 044316 (2003).
\bibitem{Dud04}	J. Dudek et al., 
		{\it Eur. Phys. J.} \textbf {A20}, 15 (2004).
\bibitem{Dub04} N. Dubray, J. Dudek and A. Maj, 
                Proceedings of this Conference.	 
\bibitem{Lach}	M. Lach et al.,
		{\it Eur. Phys. J.} \textbf {A16}, 309 (2003).		      
\bibitem{Beck}  F.A. Beck, 
               	{\it Prog. Part. Nucl. Phys.} \textbf {28}, 443 (1992).
\bibitem{Simps} J. Simpson, 
               	{\it Z. Phys.} \textbf {A358}, 39 (1997).
\bibitem{Maj94} A.~Maj et al., 
               	{\it Nucl. Phys.} \textbf {A571}, 185 (1994).
\bibitem{Her} 	M.G.~Herman, T.M. Cormier and M. Satteson, 
		{\it Phys. Lett.} \textbf {203B}, 29 (1988).
\bibitem{Puhl} 	F. P\"uhlhofer, 
		{\it Nucl. Phys.} \textbf{A280}, 267 (1977).	       
\bibitem{Kic93} M.~Kici\'nska-Habior et al., 
               	{\it Phys. Lett.} \textbf {B308}, 225 (1993).
\bibitem{Ormand} W.E. Ormand, P.F. Bortignon and R.A. Broglia, 
               	{\it Nucl. Phys.} \textbf {A618}, 20 (1997).
\bibitem{JJG} 	J.J. Gaardh{\o}je, 
		{\it Annu. Rev. Nucl. Part. Sci.} \textbf {42}, 483 (1992).
\bibitem{Nee82} K. Neerg{\aa}rd, 
               	{\it Phys. Lett.} \textbf {110B}, 7 (1982).
\bibitem{Gall}  M. Gallardo et al., 
		{\it Nucl. Phys.} \textbf {A443} 415 (1985).
\bibitem{Doss}  T. D{\o}ssing, private communication.
\bibitem{Brekiesz} M. Brekiesz et al., Proceedings of this Conference.             
\bibitem{Benzoni} G. Benzoni et al., 
		{\it Phys. Lett.} \textbf {540B}, 199 (2002).               
\bibitem{Herskind} B. Herskind et al., 
               {\it Phys. Rev. Lett.} \textbf {59}, 2416 (1987).
\bibitem{Dudek} J. Dudek, N. Schunck and N. Dubray, Proceedings of this Conference.

\end{thebibliography}
\end{document}